# Phytoscale Transport Physics: Insights into Xylem Flow Homeostasis and Drought Stress

Jinmay Kalita[1], Sumit Kumar Mehta[1], Suraj Panja[1], Pranab Kumar Mondal[1,2,]

[1]Microfluidics and Microscale Transport Processes Laboratory, Department of Mechanical Engineering, Indian Institute of Technology Guwahati, Guwahati-781039, Assam, India

[2]School of Agro and Rural Technology, Indian Institute of Technology Guwahati, Guwahati- 781039, Assam, India

[*]Corresponding author Email: pranabm@iitg.ac.in, mail2pranab@gmail.com  (P. K. Mondal)

**Abstract**

We investigate the flow dynamics of nutrient solution through the xylem vessels of *Brassica juncea* (Indian mustard) under drought stress. To this end, we perform experiments to obtain morphological traits of xylem vessels under drought-stressed conditions, and develop a mathematical framework to model the underlying flow through the xylem, considering several features relevant to the plant system. Performing experiments using state-of-the-art instruments, we measure the morphology of xylem vessels, physicochemical and mechanical properties of xylem walls under drought-stressed conditions. Our experimental results unveil that drought reduces both xylem diameter and pit aperture size, implicating hydraulic adaptations of plants to drought stress. We find that the reduced cellulose content in drought-stressed xylem vessels lowers the zeta potential and decreases elasticity of the vascular region. Additionally, drought stress alters metabolite activity, increases reactive oxygen species, reduces chlorophyll content, and limits the uptake of essential metallic nutrients. Besides, we perform three-dimensional numerical simulations to evaluate local flow field, mechanical stress, hydraulic conductivity, and radial transport efficiency of xylem vessels under drought-stressed conditions. Simulated results reveal that under drought conditions, resistance to axial flow through xylem vessels increases significantly, which in turn, promotes radial transport of nutrients, allowing plants to survive even in drought stress. We show that the radial flow efficiency of xylem vessels becomes notably higher under drought stress than in well-watered control plants. Overall, results of this endeavor provide new insights into how geometric adaptations of xylem vessels modify flow behavior under water-deficient conditions, enhancing the plant's ability to survive environmental stress.

**Introduction**

Unlike conventional engineering systems, fluid transport within plant tissues, particularly xylem vessels, is highly complex in nature owing to their reticulated network (Kim *et al.* 2014). It is non-intuitive to expect that xylem conduits are able to transport both micro and macro ionic nutrients alongside water from roots to shoots (Barceló-Anguiano *et al.* 2021; Gersony *et al.* 2020) without the aid of any mechanical pump (Webber & Worster 2023). Efforts undertaken by the researchers', so far, have revealed that this upward movement of ionic liquid is primarily actuated by a pressure gradient arising from two natural driving forces: osmotic pressure generated in the roots and the transpiration pull exerted by the leaves (De Boer & Volkov 2003). Intricate experiments further indicate that the porous pit membranes with hole structures, commonly known as pit holes, interconnecting adjacent xylem vessels serve as crucial gateways for the radial transport of nutrient-rich solutions (Choat *et al.* 2008; Park *et al.* 2018; Zhang *et al.* 2023). Porous pit membrane is typically composed of cellulose, intrinsically embedded with a negative charge due to the presence of carboxylic group therein, establish an electrohydrodynamic interplay between the ionic solution and the vessel wall. The underlying interfacial electrohydrodynamics leads to the generation of streaming potentials within the xylem (Hao *et al.* 2021; Kalita *et al.* 2025; Pereira *et al.* 2018). Quite fascinatingly, these intrinsic biophysical features of xylem transport undergo significant alterations when the plants experience biotic or abiotic stresses, leading to reformation of underlying dynamical aspects of water and nutrient transport through the plant as a whole (Dahro *et al.* 2023).

Drought-induced stress is one of the major abiotic stresses that the plants encounter primarily due to non-availability of water in the rhizosphere. The scarcity of water restricts the transport of nutrient solution through xylem, resulting in reduced crop yields and jeopardized food security (Al-Taey & Hussain 2023; Qiao *et al.* 2024). The reduced ionic absorption by the plant roots under drought stress leads to the alteration of both internal and external structures of the plant significantly (Akram *et al.* 2020; Dahro *et al.* 2023; El-Ramady *et al.* 2021; Ghafoor *et al.* 2019). Thus, the hydraulic conductivity of the xylem vessels in drought conditions depends on the anatomical variation influenced by water stress and the drought intensity (Li *et al.* 2024). It is reported that the vestured pits available in the xylem vessel walls of angiosperms offer hydraulic safety to the plants during drought (Journal *et al.* 2012). However, plants subjected to prolonged drought stress face up several adverse situations, including mortality due to carbon starvation, hydraulic failure, and pathogens attacking (Camarero 2021; Drake-schultheis 2020).

Due to the intricate structures and micro-size configuration of xylem vessels, the local variation of the flow field has not yet systematically articulated, even though the experimental results have shed light on global parameters like the averaged pressure gradient, averaged velocity, and averaged ionic concentrations associated with the flow through xylem vessel (Brodersen *et al.* 2019). Paying adequate attention to this aspect, researchers from multiple disciplines have been actively engaged in developing mathematical models essentially to understand the fundamental aspects triggering flow dynamics in xylem (de Araujo *et al.* 2021; Lee & Voit 2010; Payvandi *et al.* 2014; Thamm *et al.* 2019; Walker *et al.* 2024; Zhang *et al.* 2024; Hölttä *et al.* 2006). In this context, computational fluid dynamics (CFD) has emerged as an efficient tool for probing into the flow actuation mechanism inside the xylem (Schulte 2012; Xu *et al.* 2021, 2022), aligning closely with botanical observations. Researchers have, so far, attempted to apply one-, two-, and three-dimensional numerical models to investigate the underlying flow dynamics in the xylem vessel. Despite extensive efforts undertaken by the researchers across the scientific community (de Araujo *et al.* 2021; Lee & Voit 2010; Payvandi *et al.* 2014; Thamm *et al.* 2019; Walker *et al.* 2024; Zhang *et al.* 2024; Hölttä *et al.* 2006), challenges are still persisting to develop an exact mathematical framework that precisely replicates the transport phenomena due to the axio-radial flow in the xylem vessel and its complex morphological structures. Nonetheless, using a simplified one-dimensional mathematical model for the wheat plant, the transport of phosphate and water in the xylem vessels has been investigated (Roose 2014). Effort has also been taken to reconnoiter the flow inside the xylem vessels employing two-dimensional numerical model with annular thickening of vessel walls (Roth 1996). The study reported that the height and spacing of two adjacent rings of the wall with annular thickening have a significant impact on the flow characteristics in xylem vessels (Roth 1996). In contrast to the smooth vessel model, the three-dimensional modelling of xylem vessel considering circular cross-section, which mimics xylem vessels with secondary thickening, indicated the lowest flow resistance. This observation is suggestive of that consideration of a simple xylem model results in underprediction of transport efficiency. Another study available in this paradigm, which is consistent with three-dimensional numerical simulations, revealed that increasing the diameter of xylem vessels resulted in a faster flow velocity while reducing pressure drop and flow resistance coefficient (Chen *et al.* 2015). However, adding a scalariform perforated plate to the xylem model, which is practically relevant as well, increased the overall pressure drop and flow resistance while maintaining a constant average flow velocity as the plate's hole count increased (Ai QingLin *et al.* 2011). Despite several models, discussed above, have been able to provide a basic understanding of the xylem flow dynamics, a poor correlation of the aforementioned models to the experimental data restricts these frameworks to reproduce multiphysical aspects of xylem transport accurately.

The foregoing discussion underscores the crucial role of xylem physiology in plants' survival under drought-stressed condition, and reviews of existing models largely employed to compute the flow field in xylem vessels. Nevertheless, a state-of-the-art mathematical framework for xylem transport that takes drought stress into account, while aligning with experimental findings remains a critical gap until this endeavor. To this end, we perform experiments to obtain morphological traits of xylem vessel and develop a mathematical framework to model xylem flow using experimentally measured parameters. The primary objective of this study is to experimentally investigate the morphological variations in xylem vessels under both normal and drought-stressed conditions. This enables us to provide accurate geometrical inputs to modelling framework developed in this endeavor. By integrating these experimentally measured findings with simulations, we visualize the local flow field developed in the xylem vessels and perform a comprehensive analysis of mechanical stresses in correlation to biological responses under control and drought conditions. In summary, this research combines experimental investigations of the physio-biochemical responses of plants to drought stress with full-scale numerical simulations of xylem flow. We believe that the proposed integrated approach offers valuable insights into the behavior of xylem hydraulics under drought conditions.

**Problem formulation**

In the present study, we first experimentally investigate the morphological traits, biochemical properties, and electrokinetic variations in the xylem vessels of *Brassica juncea* under both control and drought conditions. To this end, we conducted experiments for Scanning Electron Microscope (SEM) imaging, zeta potential measurement, estimation of streaming potential, inductively coupled plasma mass spectrometry (ICP-MS), laser Raman spectroscopy, atomic force microscopy (AFM), fluorescence microscopy, and leaf chlorophyll content analysis for both control and drought scenarios. We further develop a three-dimensional model of xylem vessels, consistent with experimental data, corresponding to analyze as well as understand the drought stress induced variations of flow physics inside the plants. In this context, we systematically examine axial and radial flow velocities along with flow loading, net throughput, flow-induced mechanical stresses, hydraulic efficiency as well as radial flow efficiency of xylem vessels under both control and drought-stressed conditions. We undertake this effort to analyze the inpacts of environmental conditions, which are varying dynamically across the glove, on the transport efficacy of xylem vessels.

**Experimental approach and analysis**

*Materials and methods*

Using the control condition as a reference, we perofmr a series of experiments to investigate the morphological, biochemical, electrohydrodynamical, and structural responses of xylem vessels to drought stress. As mentioned before, we consider *Brassica juncea* (Indian mustard) for this study, as it is the third-largest source of edible oil globally (Indira *et al.* 2021).

*Plant species and sample preparation*

The plant considered in the present study is 45-day-old *B. juncea*, cultivated in the greenhouse of the Indian Institute of Technology Guwahati during the month of January. Throughout the growth period, the average ambient temperature and relative humidity were recorded as $18.4 \pm 5$ °C and $77.8 \pm 3$ % respectively. We collected seeds for the experiments from the ICAR–Indian Agricultural Research Institute (IARI), Regional Station in Karnal, Haryana, India. To prevent any potential pathogenic contamination, we went for surface sterilization of the seeds using a 4% (w/v) sodium hypochlorite solution. Seeds were then thoroughly rinsed 3–4 times with autoclaved double-distilled water to remove any residual chemicals before being sown in pots, each containing 1.5 kg of soil. The seedlings were initially grown in the greenhouse for 38 days under identical conditions, including water supply, and maintaining standard agricultural protocols (Huynh *et al.* 2023). After this period, drought stress is induced in a test set of plants for 7 days employing water withdrawal method, while control groups were nurtured with regular watering. Subsequently, we conducted experiments using both the control and drought-stressed plants, with at least three replications of each case.

*Scanning Electron Microscopy (SEM) imaging*

We undertook an effort to capture SEM imaging of xylem vessels morphology for both control and drought-stressed plants employing the protocol mentioned by Nikara *et al.* (2020). For the sake of completeness, we here briefly mention the protocols as follows: 1 cm long stem segment was cut at a distance 10 cm from the root-shoot junction. The lateral and longitudinal sections of the stem segment were preserved in 2.5 % glutaraldehyde (SRL) in 0.1 M sodium phosphate buffer for 12 h at 4° C, followed by the dehydration with 25%, 50%, 70%, 90% and 100% ethanol for 15 mins each. The samples were finally dried using hexamethyldisilazane (HDMS, Sigma-Aldrich) for overnight to obtain the morphological images. The images were captured by a Field Emission Scanning Electron Microscope (Make: Zeiss®, Model: Sigma 300).

*Measurement of zeta potential*

We measured zeta potential for the xylem vessels of control and drought-stressed plants to estimate the surface charge inherited by the vessels. A longitudinal section, mostly comprising of xylem alongside a few cortical cells, as confirmed by Scanning Electron Microscopy (SEM) images was extracted from the stem segments. The section was finely crushed to powder in a Morter, which is further sonicated with 5 mL deionized (DI) water in a conical flask until a colloidal solution is formed. The Zeta potential of this colloidal solution was then measured using Anton Paar® Dynamic Light Scattering (DLS) instrument with model Litesizer 500 (Behera *et al.* 2022).

*Laser Raman analysis*

We measured the levels of metabolites both in control and drought-stressed plants using a Laser-based Micro Raman system (HORIBA Jobin Yvon®, Model: LabRAM HR) (Panja *et al.* 2024). We collected fresh longitudinal sections of stem containing xylem vessels from both control and drought-treated plants for the analysis on which a laser beam was focused through a 50x magnification lens.

*Inductively coupled plasma-mass spectrometry (ICP-MS) analysis*

The quantitative variations in various ionic nutrients between control and drought-stressed plants were determined using ICP-MS with model Agilent® 7850 (Panja *et al.* 2024). Below we outline the experimental procedure. We extracted 100 mg plant dust from the stem segments of control and drought-treated plants. The samples were then digested with hydrogen peroxide ($H_2O_2$) and Nitric acid ($HNO_3$) in 9:1 ratio. The digested samples were then diluted with DI water to 50 mL volume. Following this, 10 mL of the prepared samples were filtered for ICP-MS analysis to detect the concentration of the ionic nutrients.

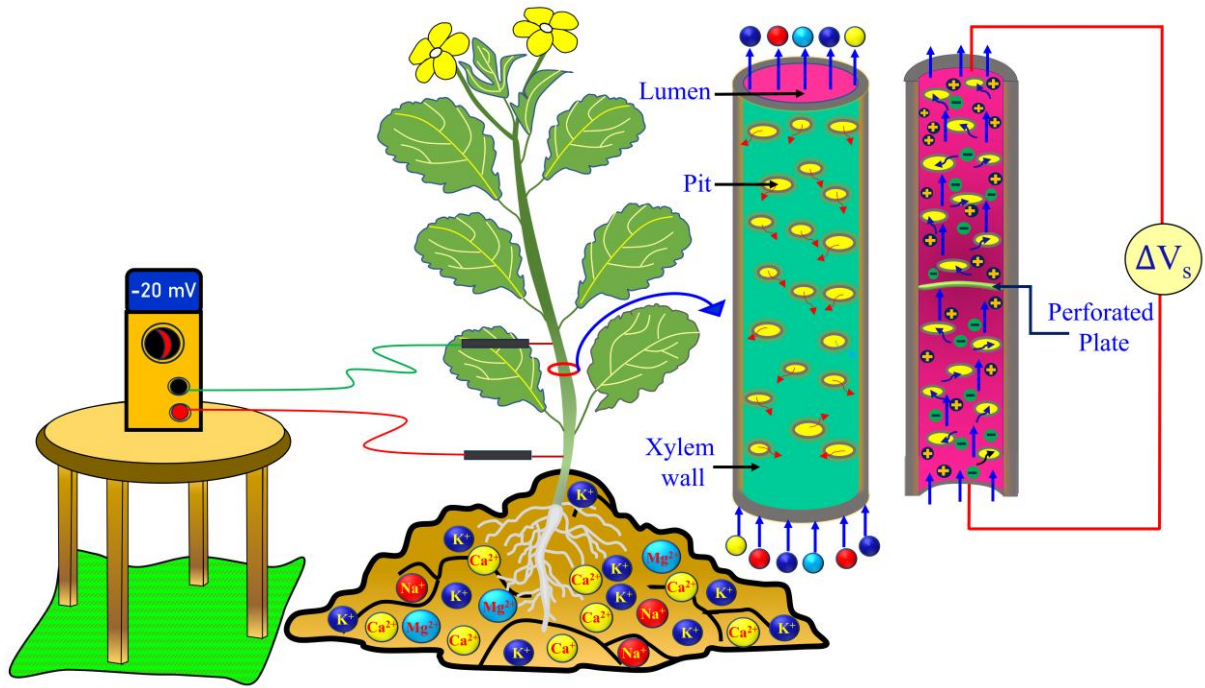

**Figure 1.** Representative image of streaming potential measurement procedure of xylem vessels in which streaming potential is induced due to the interaction between ionic nutrient solution and the charged xylem walls.

*Measurement of streaming potential*

The streaming potential of *B. juncea* was measured along the height of plant, as illustrated schematically on the left side of the figure 1, to determine the pressure gradient induced due to flow through xylem vessel. Note that upward flow of ionic liquid through xylem vessels is primarily actuated by a pressure gradient arising from two natural driving forces: osmotic pressure generated in the roots and the transpiration pull exerted by the leaves. This potential is induced in plants because of streaming of the counter ions, known as streaming potential, resulting from an interplay between the inherent ionic nutrient flow through xylem vessels and the charged xylem walls (cf. right side of figure 1). Interested readers may refer to the seminal works on this part (Chang and Leslie 2010). The apparatuses used during the measurement of streaming potential were a FLUKE® 179 True-RMS digital multimeter and two brass electrodes. One of the electrodes was inserted 2 mm deep into the plant at the junction of soil and the plant's shoot acting as reference electrode, while the distance of insertion of the other electrode from the reference electrode was varied along the stem at 5 cm interval up to 25 cm. The streaming potential was measured by connecting the positive and negative terminals of the multimeter to the exposed ends of the two electrodes. The readings were taken every 5 minutes over a 45 minutes' time period for both control and drought-induced plants.

*Atomic Force Microscope (AFM)*

We performed AFM analysis of the longitudinal sections of the xylem of both control and drought-induced plants to measure the elastic property of xylem vessel. For this part, we prepared both samples following the same protocols as employed for zeta potential measurement. The instrument used for the analysis is MFP-3D-BIO-Asylum Research-Oxford Instruments equipped with a conical indenter (Panja *et al.* 2024). The spring constant ($k$), Young's modulus and Poisson's ratio of the indenter were specified as 2 N/m, 169 GPa and 0.22 respectively. Now, the Young's modulus ($E_x$) and Poisson's ratio ($v_x$) of the xylem are calculated from the following relationship:

$$\frac{1}{E^*} = \frac{1-v_t^2}{E_t} + \frac{1-v_x^2}{E_x} \tag{1}$$

Where $E^*$ is the equivalent Young's modulus which can be assessed using following expression incorporating the contact force ($F_c$), cone's half angle ($\alpha$) and indentation depth ($h$)

$$F = \frac{2}{\pi}\tan(\alpha)E^*h^2 \tag{2}$$

*Fluorescent microscopy images*

The generation of reactive oxygen species (ROS) in the plants under both control and drought conditions was assessed using a ROS-sensitive fluorescent dye [2,7-dichlorofluorescein diacetate ($H_2$DCF-DA), SRL)] following a revised protocol adapted from Awasthi *et al.* (2019). The longitudinal sections of the stem, containing xylem vessels from both control and drought-stresses plants were immersed in a 10 mM Tris-HCl buffer, prepared from 10 μM $H_2$DCF-DA fluorescent probe. The pH of the buffer was adjusted to 7.2. After 20 minutes of incubation in the dark to prevent photobleaching, the samples were washed with Tris-HCl buffer. Subsequently, ROS accumulation was visualized under a fluorescence microscope (BestScope BS-7000A).

*Quantitative analysis of chlorophyll A content*

The chlorophyll A content of both control and drought-stressed plants were measured using a CI-710 s SpectraVue Leaf Spectrometer. Prior to the measurements, the spectrometer was carefully calibrated. Subsequently, the intact leaves of the control and drought plants were placed inside the leaf clip of the instrument to record the data indicating Chlorophyll A content. The experiment was repeated for three leaves of both control and drought-stressed plants.

*Description of morphological traits of xylem vessels*

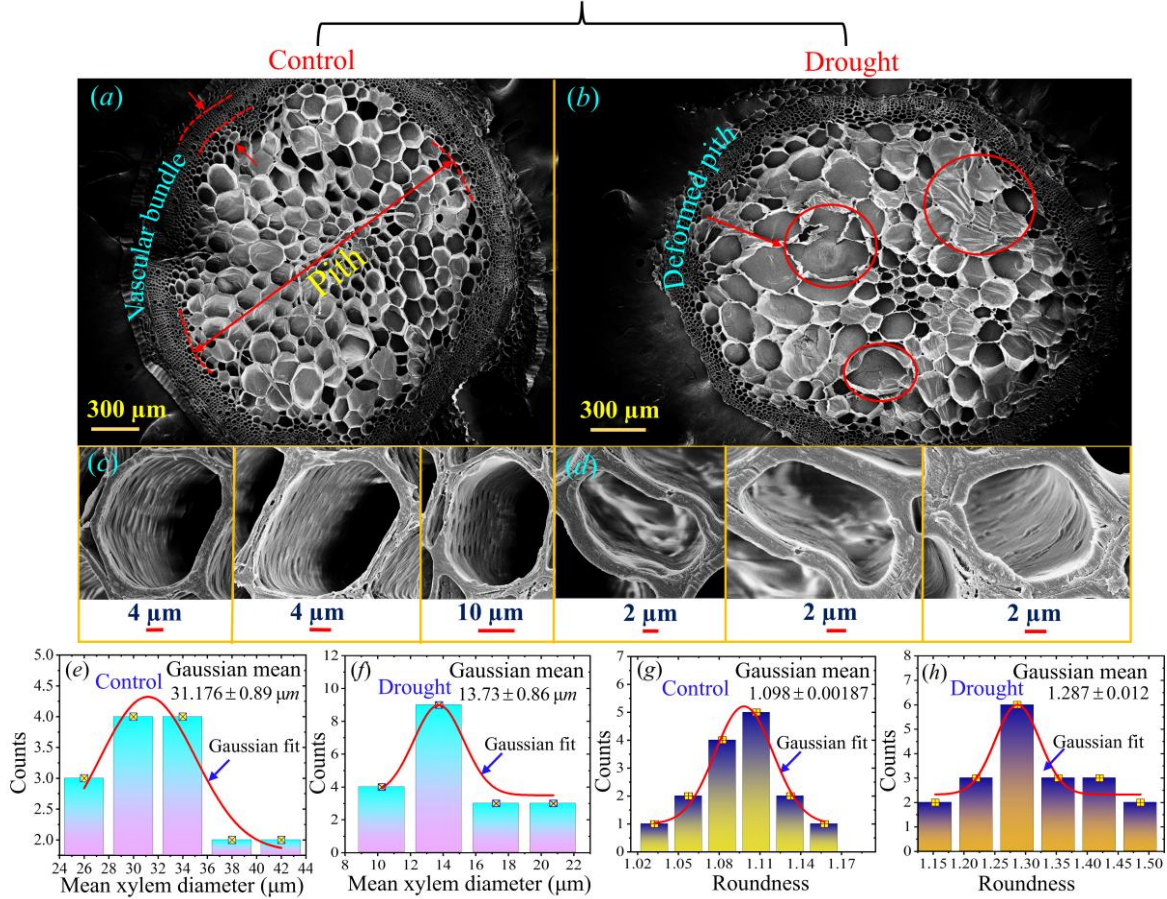

**Figure 2.** (*a*) Scanning electron microscope images of anatomical sections of *Brassica juncea* highlighting vascular bundle and piths in control and drought conditions. The piths are observed to deform during drought-induced stress. (*b*) Zoomed-in view of pitted xylem vessels found in vascular bundle of *B. juncea* under (*c*) control and (*d*) drought environments. Gaussian distribution of mean xylem vessel diameter for (*e*) control and (*f*) drought cases. Gaussian distribution of the roundness parameter of xylem vessels under (*g*) control and (*h*) drought conditions. While evaluating the xylem diameter and roundness parameter, a total of 15 and 19 xylem vessels were analyzed in control and drought-treated, respectively.

To investigate the internal morphological changes in *B. juncea* under drought conditions, we represent scanning electron microscope (SEM) images of lateral anatomical sections from both control and drought-affected plants in figures 2(*a*)-(*b*). The images reveal that the morphological structure of the pith is noticeably deformed in plants exposed to drought stress. This disruption of pith parenchyma cells is likely to be triggered by increased levels of abscisic acid (ABA), which result from the enzymatic oxidation of carotenoids following their depletion (Higgins *et al.* 2022; Pressman *et al.* 1983). A more detailed discussion on the metabolic alterations, induced by drought stress, is provided in the subsequent phase of the current study. Pith deformation syndrome eventually leads to the development of hollow stems in drought-stressed plants, impeding the growth (Pressman *et al.* 1983).

We further examine the cross-sectional variation in xylem vessels between control and drought-stressed plants, and the corresponding SEM images are presented in figures 2(*c*)-(*d*). The cross-sectional view of the xylem vessels is observed to be approximately 'round-shape' in control plants. In contrast, a deviation from the ideal circular shape is noticed in drought-stressed plants. Under drought stress, a significant reduction in xylem lumen diameter is observed compared to the control condition. This structural adaptation helps plants to uphold hydraulic safety under water-deficit conditions, thereby enhancing their ability to the drought resistance (Liu *et al.* 2023a). To quantify the size of xylem vessels in both control and drought-stressed plants, we estimate the Gaussian distribution of mean xylem diameters in figures 2(*e*) and 2(*f*), respectively. We employ ImageJ software to extract the geometrical information of xylem vessels from the SEM images. The analysis reveals that the Gaussian mean xylem diameters are 31.176 ± 0.89 μm and 13.73 ± 0.86 μm, based on measurements from 15 and 19 xylem vessels in control and drought-treated plants, respectively.

Due to the apparent irregular shapes of xylem vessels in the SEM image, we take an effort to enumerate the degree of roundness in terms of roundness parameter (£) of the xylem cross-sections using the following expression (Hillabrand *et al.* 2019):

$$£ = \frac{Perimeter^2}{4\pi \times Area} \tag{3}$$

Note that £ = 1 represents a perfect circle, while values of R greater than unity indicate deviations from an ideal circular shape. We present the roundness values for both control and drought-treated plants in figures 2(*g*)-(*h*). It is evident that the xylem vessels in control plants maintain an approximately circular shape with £ values remaining closer to 1 (=1.00187). This observation is attributed to the greater contents of cellulose, which enhance the mechanical strength resisting the compressive forces exerted by water conduction, thereby supporting cellular integrity (Crowe *et al.* 2021; Joshi *et al.* 2011). In contrast, the xylem vessels of drought-stressed plants deviate significantly from the ideal circular shape with £ = 1.287 (> 1) as compared to that observed in control plants. This irregularity arises from a reduced cellulose content in the vessel walls under drought conditions. As a result, the xylem walls become weaker and less rigid under drought stress (Froux *et al.* 2004; Taylor *et al.* 1999). Consequently, the structural integrity of the xylem vessels is impaired under drought environments.

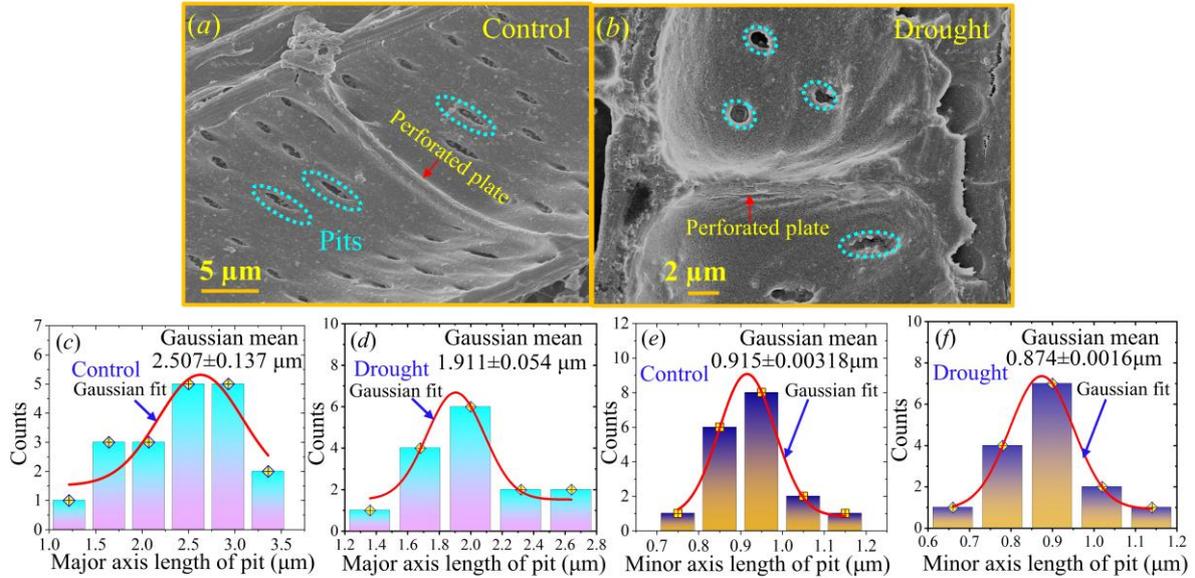

**Figure 3.** Scanning electron microscope images of xylem wall pits of *B. juncea* for (*a*) control and (*b*) drought cases. Gaussian distribution of major and minor axis length of the elliptical pits for (*c*), (*e*) control and (*d*), (*f*) drought conditions, respectively. The analysis included 19 xylem wall pits from control and 15 from drought-treated plants.

As the transport through xylem vessels is largely influenced by inherent pit structures, we further investigate the morphological variation of xylem wall pits under control and drought conditions. Accordingly, we present the SEM images of pit structures for both the water availability conditions in figures 3(*a*)-(*b*). The xylem pits appear to be approximately elliptical in shape for both cases. Interestingly, the drought-affected xylem vessels exhibit noticeably smaller pits compared to those in the control group. The reduction in pit aperture size under drought conditions is essentially a fundamental hydraulic adaptation to resist the loss of water. This structural adjustment, triggered by water stress, reduces the risk of embolism formation that can disrupt water transport in the xylem (Liu *et al.* 2023a). Further, we measure the major and minor axis lengths of the elliptical-shaped xylem pit apertures for both control and drought-stressed plants, as shown in figures 3(*c*)–(*f*). The Gaussian mean major axis lengths for the control and drought conditions are estimated to be $2.507 \pm 0.137$ μm (cf. figure 3c) and $1.911 \pm 0.054$ μm (figure 3d), respectively, while the corresponding Gaussian mean minor axis lengths are calculated as $0.915 \pm 0.00318$ μm (figure 3e) and $0.874 \pm 0.0016$ μm (figure 3f), respectively.

*Description of biochemical and biomechanical traits of xylem vessels*

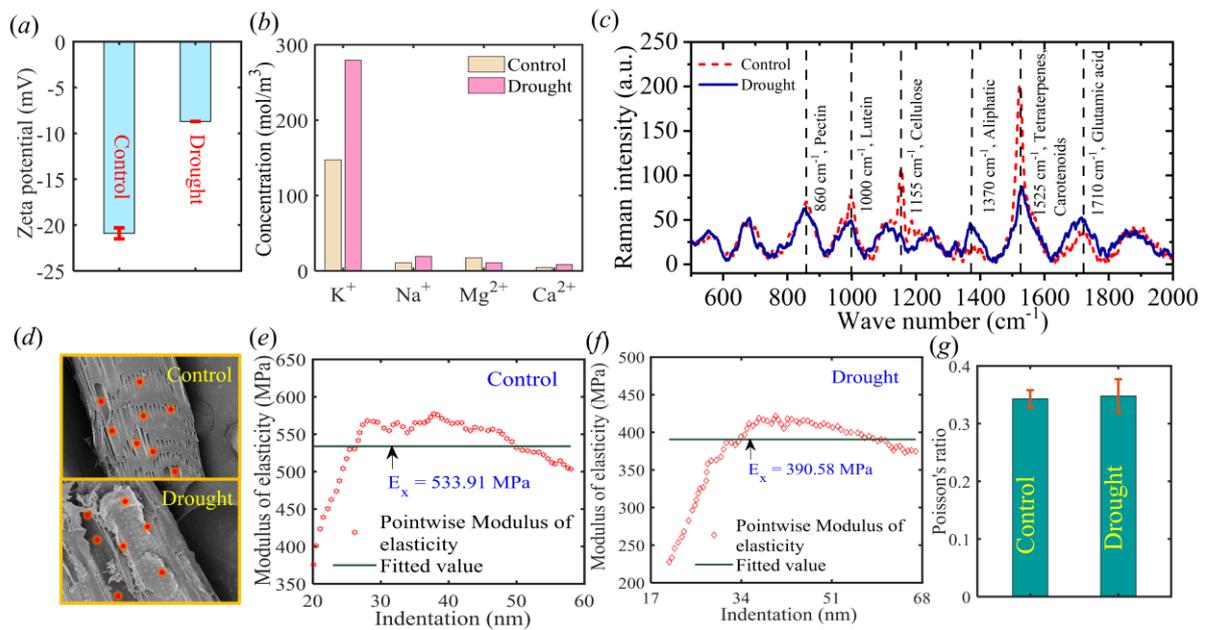

**Figure 4.** Variation in (*a*) vascular region zeta potential and (*b*) vital ionic nutrient concentration obtained from Inductively Coupled Plasma Mass Spectrometry (ICP-MS) analysis (*c*) Raman intensity of different metabolites during control and drought conditions. (*d*) Representative experimental samples of vascular bundles of *B. juncea* containing mostly xylem vessels along with other cellular structures employed to measure mechanical properties of xylem under control and drought conditions using AFM. Point-wise variations of modulus of elasticity alongside mean modulus of elasticity under (*e*) control and (*f*) drought conditions. (*g*) Variations in Poisson's ratio for control and drought-treated xylem vessels.

We now look into the variations in biochemical characteristics of the stem vascular region under drought-stressed condition. It may be mentioned here that the biochemical properties of stem vascular region are very much involved with the underlying nutrient transport through the xylem vessels. This is done by examining zeta potential (cf. figure 4*a*), nutrient concentration (cf. figure 4*b*) and Raman intensity (cf. figure 4*c*) of the vascular region. It is well established that xylem walls possess a negative surface charge, primarily due to the presence of carboxylic groups in cellulose, which is a key structural component of vessel. Under drought conditions, a reduction in the negative zeta potential of the vascular region is observed, as witnessed in figure 4(*a*). This diminution in surface charge is associated with a decrease in cellulose content in drought-stressed plants, which will be discussed in detail in the context of Raman analysis presented in figure 4(*c*).

In order to delve deep into the biochemistry of xylem vessels, we analyze the metallic ion concentrations in stem under the control and drought-stressed plants using ICP-MS (cf. figure 4*b*). The data show a significant increase in potassium ($K^+$) levels under drought conditions. This rise in K$^+$ is linked to the preservation of water balance through osmotic pressure adjustment, known as water homeostasis (Martineau *et al.* 2017; Srivastava *et al.* 2020). Higher potassium accumulation during drought stress also contributes to oxidative damage by forming reactive oxidative species during photosynthesis (Seleiman *et al.* 2021). Similarly, an increase in sodium (Na$^+$) concentration is observed in drought-affected plants. This is probably due to the accumulation of sodium ions in the upper soil layers as the soil dries out (Ma *et al.* 2020). The elevated sodium levels contribute to osmotic stress and ion toxicity in plants under drought conditions (Ma *et al.* 2020). In contrast, magnesium (Mg$^{2+}$) levels are reduced during drought-induced stress. This decrease is attributed to impaired uptake by the roots, caused by reduced water movement and ionic diffusion in dry soil (Senbayram *et al.* 2015). As magnesium is a critical component of chlorophyll, its deficiency leads to a decrease in chlorophyll content, resulting in chlorosis (Ahanger *et al.* 2016). Note that a deficit in chlorophyll content with a reduction in magnesium is witnessed in figure 11(b), shown in Appendix A. Moreover, low magnesium levels negatively impact the synthesis of carotenoids and lutein (cf. figure 4c) by downregulating key carotenoid biosynthesis genes (CitPSY, CitLCYe, and CitPDS). Consequently, photoassimilates accumulate in the leaves are not effectively transported to other plant parts, leading to chloroplast deformation and reduced CO$_2$ diffusion from leaves (Liu *et al.* 2023b; Senbayram *et al.* 2015). This imbalance promotes the generation of reactive oxygen species (ROS) in plants under drought stress (cf. figure 11(a) of Appendix A). A slight increase in calcium (Ca$^{2+}$) concentration is also noted in drought-stressed plants, possibly due to elevated glutamic acid levels (cf. figure 4*c*). The rise in calcium signaling results in the accumulation of abscisic acid (ABA) that enhances drought tolerance (Zuo *et al.* 2025). Changes in calcium levels activate drought-responsive genes such as RD29A/B and KIN1/2, which produce protective proteins (Gupta *et al.* 2023). Additionally, it is involved in the function of plasma membrane ATPase - an enzyme crucial for restoring nutrient balance following stress-induced membrane damage (Ahanger *et al.* 2016).

It is important to mention here that the change in water availability conditions leads to a significant variation in plant metabolism. Considering this aspect, we present the Raman intensity peaks for some important metabolites, namely glutamic acid, carotenoids, tetraterpenes, aliphatic compounds, cellulose, lutein, and pectin under both control and drought conditions (cf. figure 4*c*). Notably, the peak intensity of glutamic acid is higher in drought-stressed plants. Elevated glutamic acid levels are known to promote calcium signaling, which in turn, activates salicylic acid synthesis (Qiu *et al.* 2020). This signaling pathway promotes protective proline accumulation under drought conditions, thereby strengthening the plant's defense mechanisms (Qiu *et al.* 2020). In contrast, the Raman intensity of carotenoids drops significantly under drought stress. This reduction impairs the plant's photosynthetic efficiency (cf. Appendix A) and its ability to neutralize free radicals, as carotenoids play a crucial role in photoprotection (Choudhury & Behera 2001). Additionally, the deterioration in carotenoid content under drought leads to increased enzymatic oxidation and the accumulation of abscisic acid (ABA). Albeit ABA enhances plant's ability to withstand drought resistance, it contributes to the degradation of pith parenchyma cells (Higgins *et al.* 2022; Pressman *et al.* 1983). Furthermore, the accumulation of aliphatic compounds is enhanced in drought-stressed plants. These compounds improve drought resilience by reducing water loss (Macková *et al.* 2013). However, drought stress also leads to a decrease in the contents of lutein, pectin, and cellulose. Reduced cellulose content results in both a lower zeta potential of vascular region (cf. figure 4*a*) and a decreased modulus of elasticity (discussed in upcoming paragragraph) of xylem vessels. A decrease in lutein levels compromises both photosynthetic efficiency and photoprotection, leading to elevated production of reactive oxygen species (ROS) (cf. Appendix A), which in turn, impairs overall plant growth. The reduction in pectin content under drought conditions further intensifies the attenuation of the water retention capacity and hampers cell wall integrity (cf. figure 2*b*). In summary, these cumulative changes in key metabolites under drought stress significantly alter both the external morphology and internal electrohydrodynamic properties of xylem vessels, ultimately affecting the plant's overall ability to withstand water deficit conditions.

    We further investigate the variation in structural properties of the xylem vessels in control and drought cases, as these are important input parameters for modeling framework to understand the flow dynamics through soft xylem vessels. In an effort to measure the structural properties of soft xylem vessel through experiments, we prepared samples as represented in figure 4(*d*). Accordingly, we measured the modulus of elasticity and Poisson's ratio of the xylem vessel for both cases, as presented in figures 4(*e*)-(*g*). Quite intuitively, the elastic modulus of xylem vessels in drought plants is found to be lower compared to drought conditions. This is because of the reduced content of structural metabolites such as cellulose and pectin under drought (cf. figure 4c), which in essence compromises the stiffness. The lower elastic modulus in drought conditions also contributes to a greater lateral strain in the xylem vessels, leading to a slightly higher Poisson's ratio value compared to control plants. This increased lateral deformation causes the cross-section of the drought-stressed xylem vessels to deviate from its ideal circular shape and a greater value of the roundness parameter ($\mathcal{E} > 1$) (c.f. figure 2*b*).

*Description of electrohydrodyanmic traits of xylem vessels*

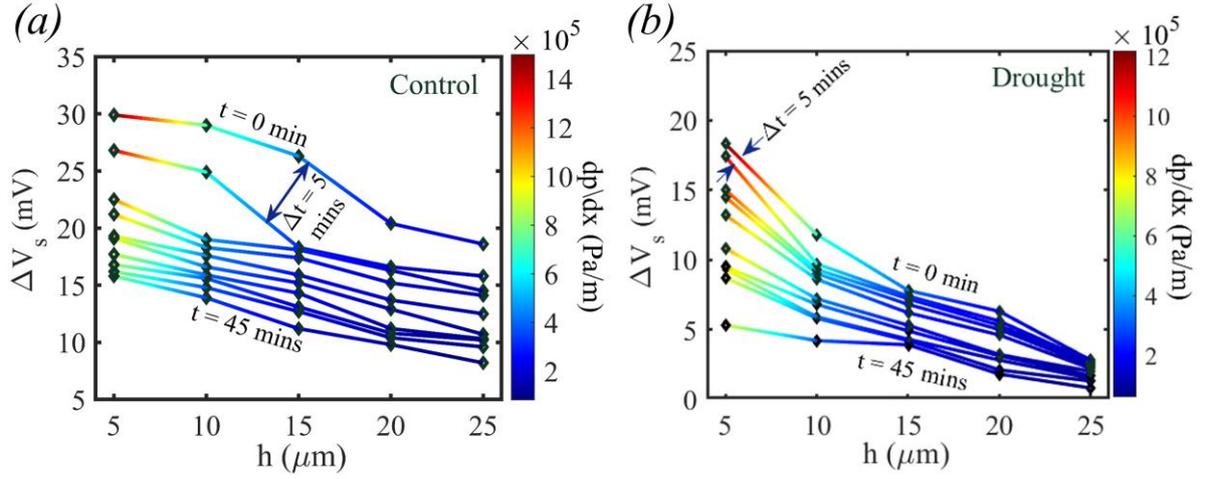

**Figure 5**. Variation in streaming potential of xylem vessels at different heights of the plant along with pressure gradient (color-coded) at varying time intervals in (*a*) control and (*b*) drought conditions.

Furthermore, to understand electro-hydrodynamics in xylem vessels, we analyze the temporal variation of streaming potential along with pressure gradient (color-coded) under both control and drought conditions. We have outlined the protocols followed to measure the streaming potential in Section 3.1.6 of this article. As shown in figures 5(*a*)–(*b*), the streaming potential is consistently higher in control plants compared to drought-stressed ones. This observation is primarily attributed to the higher zeta potential of the vascular region of plants under control condition (cf. figure 5*a*), which in turn, results in a higher magnitude of reversed electroosmotic mobility because of the induced streaming electric field. Consequently, more counter-ions are displaced by the upward flow in xylem, generating a greater electrical potential difference (or streaming potential difference). In contrast, a lower zeta potential magnitude and restricted axial movement of ions due to the reduced flow strength caused by dehydration in drought conditions decrease the streaming potential difference in the xylem vessels. Remarkably, a gradual decline in streaming potential is observed along the height of the plant (cf. figure 5*b*). This decrease can be attributed to two main factors. First, radial leakage of ionic nutrients through the xylem pitted walls leads to the reduced counter-ion accumulation at upper locations in the plant. Another factor for this decline in streaming potential could be the increased axial resistance due to the longer hydraulic pathway (Liu *et al.* 2019) and the narrowing of xylem vessel diameter from root to shoot, which further diminishes the underlying flow strength (de Moraes *et al.* 2022). From experiments, we observe a decrease in streaming potential for both control and drought-stressed plants as the time progresses. The decline in streaming potential with time is likely to be attributed to the puncturing of the xylem vessel due to the insertion of the electrodes during the measurement process. This causes radial leakage of ionic liquid and weakens the force magnitude driving the flow through xylem. Knowing the value of streaming potential, we can estimate the pressure drop using the relation as follows:

$$\Delta p \approx \frac{\Delta V_s \sigma \mu}{\varepsilon_0 \varepsilon_r \zeta} \qquad (4)$$

Here, $\Delta p$, $\Delta V_s$, $\sigma$, $\mu$, $\varepsilon_0$, $\varepsilon_r$, and $\zeta$ denote the pressure drop, streaming potential, electrical conductivity, dynamic viscosity, electrical permittivity of free space, relative electrical permittivity and measured zeta potential of xylem vessels. The detail derivation of equation (3.2) is outlined in Appendix B. Note that, the resulting drop in streaming potential attenuates pressure drop in drought-stressed plant along its height, as this flow parameter is directly correlated with the streaming potential difference. Also, a reduction in pressure drop is evident over time, attributed primarily to the puncturing of xylem vessel as mentioned before. Since the pressure drop variation exhibits an asymptotic trend with time, we use the height-averaged pressure gradient at $t = 0$ in our numerical simulations for both the control and drought-stressed cases to ensure the accuracy of the results.

**Mathematical modelling**

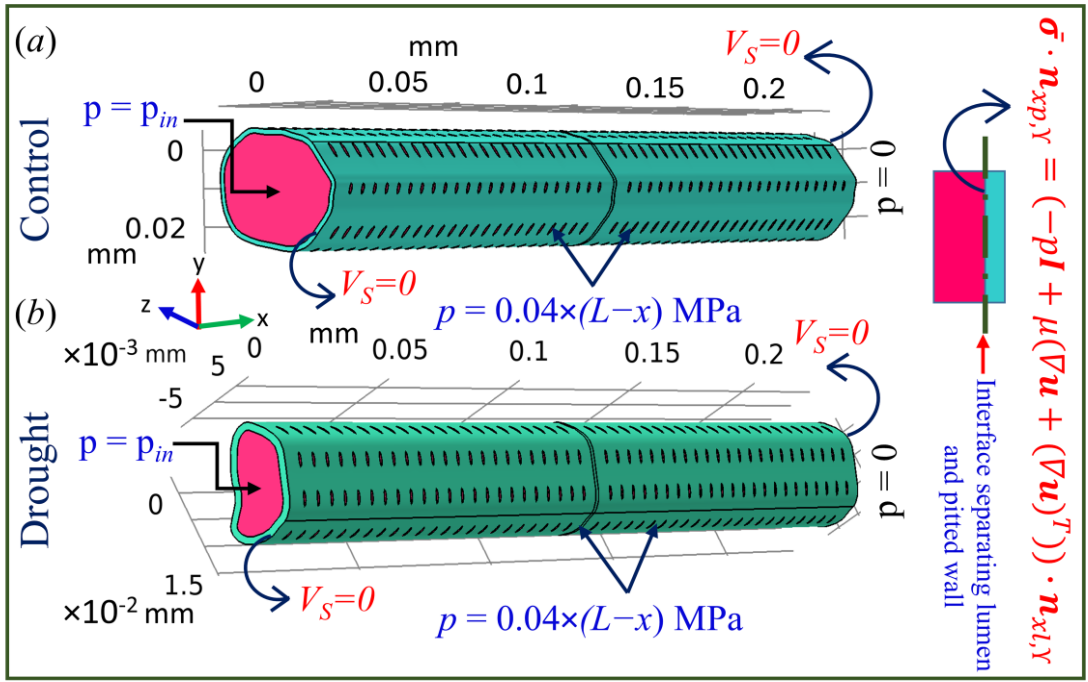

**Figure 6**. Computational domain of xylem vessel highlighting the boundary conditions imposed to solve flow and displacement fields of the xylem under control and drought conditions.

We undertake an effort to develop a modelling framework to simulate as well as analyze the influence of environmental conditions on the flow physics of xylem vessels. For the mathematical model, we consider the geometrical configuration of xylem vessels, experimentally captured through SEM imaging. Consequently, we solve the transport equations to obtain both flow field in the xylem and displacement field of xylem vessels under control and drought-stressed conditions.

*Description of flow and displacement fields under control and drought conditions*

In the current work, we consider that the flow of ionic nutrient solution within the xylem vessels is incompressible, steady and it mimics the Newtonian fluid behavior. The dimensional constraints of the xylem vessel are extracted from the SEM images using ImageJ software to develop the three-dimensional model, as illustrated in figure 6, of the vessel for numerical simulations. The flow field developed in the xylem vessel is solved employing the continuity and momentum transport equations, as given below (Mehta *et al.* 2025; Panja *et al.* 2024):

$$\nabla \cdot \boldsymbol{u} = 0 \tag{5}$$

$$\rho(\boldsymbol{u} \cdot \nabla)\boldsymbol{u} = -\nabla p + \nabla \cdot \left(\mu(\nabla \boldsymbol{u} + (\nabla \boldsymbol{u})^T)\right) \tag{6}$$

Here, the velocity vector, pressure, density and dynamic viscosity are expressed by the symbols $\boldsymbol{u}\ (\equiv u_x \hat{i} + u_y \hat{j} + u_z \hat{k})$, $p$, $\rho$ and $\mu$ respectively.

Furthermore, we employ the linear elastic model in the numerical simulation to determine the flow-induced mechanical stress inside the vessel, caused by flow loading on the lumen-wall of xylem. Note the flow through xylem vessel falls in the low Reynolds number regime (Re ≪ 1). This assumption holds good for the present case because the low Reynolds number flow in the xylem results in a negligible deformation of the vessel wall. The differential equation for structural domain of vessel wall under flow loading can be written as (Agarwal *et al.* 2024):

$$\nabla(\boldsymbol{FS})^T = 0 \tag{7}$$

Here, $\boldsymbol{F}$ and $\boldsymbol{S}$ are the displacement gradient and second-Piola-Kirchhoff stress respectively. Now, $\boldsymbol{F}$ can be determined using the displacement field ($U$) as follows:

$$\boldsymbol{F} = \boldsymbol{I} + \nabla U \tag{8}$$

$\boldsymbol{I}$ represent the identity matrix in equation (8)

Further, $\boldsymbol{S}$ can be calculated from the following expression:

$$\boldsymbol{S} = 2\mu_L \bar{\varepsilon} + \lambda_L tr(\bar{\varepsilon})\boldsymbol{I} \tag{9}$$

Here, $\bar{\varepsilon} = 0.5(F(F)^T - I)$ is Lagrange-Green strain. Note that $\lambda_L$ and $\mu_L$ are the first and second Lame parameters appeared in equation (4.5), which can be assessed in terms of Young's modulus ($E_x$) and Poisson's ratio of the xylem vessel ($\nu_x$) under control and drought-stressed scenarios as below:

$$\lambda_L = \nu_x E_x/(1 + \nu_x)(2\nu_x - 1), \mu_L = E_x/2(1 + \nu_x) \tag{10}$$

Finally, the Cauchy stress tensor for the walls of xylem can be expressed incorporating $F$ and $S$ as:

$$\bar{\sigma} = \frac{1}{J}(FS(F)^T) \tag{11}$$

$J$ denotes the Jacobian of $F$.

Consequently, the aforementioned equations are solved by employing physically justified boundary conditions, consistent with experimental evidence, as illustrated in figure 6. For equations (5) - (6): At inlet, $p = p_{in}$; at outlet: $p = 0$; at pitted wall surrounding: $p = 0.04 \times (L - x)$ MPa. Here, $p_{in}$ is the pressure at the inlet of the xylem vessels for control and drought-stressed conditions. This pressure condition is estimated using the experimentally measured pressure difference, $\Delta p = (\Delta V_s \sigma \mu/\varepsilon \zeta)$ (see Appendix B). Based on this experimentally measured pressure difference, the pressure gradient in xylem is estimated as $(\Delta V_s \sigma \mu/\varepsilon \zeta)/l$; here, gape between electrodes inserted in stem is $l$. Hence, for the xylem length ($L \sim 230$ μm from SEM image) taken for the simulations, the value of $p_{in}$ can be expressed as $(\Delta V_s \sigma \mu/\varepsilon \zeta)L/l$. For equations (7) -(11), we employ the conditions as follows: At the transverse plane of the inlet and outlet of the xylem vessels, $U = 0$ (no - displacement); at the lumen-wall interface, $\bar{\sigma} \cdot \boldsymbol{\eta}_{s,\Gamma} = \left(-pI + \mu(\nabla u + (\nabla u)^T)\right) \cdot \boldsymbol{\eta}_{f,\Gamma}$. where, $s$ and $f$ represent solid and fluid domains, respectively. Here $\boldsymbol{\eta}$ is the unit normal vector at the interface.

*Numerical analysis*

The governing equations (equations (1) to (7)) associated with the flow and displacement field of xylem vessels are numerically solved in a three-dimensional configuration (cf. figure 6), employing the boundary conditions mentioned above. The finite element framework of COMSOL Multiphysics is employed to solve the transport equations describing flow and the displacement fields of xylem vessels under both control and drought environments. We discretize the computational domain into the non-uniform mesh composed of tetrahedral elements. For the numerical implementation, linear shape function is used for velocity field and pressure, while quadratic Lagrange shape function is employed for displacement to discretize differential equations. Hence, combining all local matrices, the global matrix is generated. Subsequently, the flow field is computed using parallel direct sparse solver (PARDISO) and displacement field is solved by adopting the conjugate gradients solver.

*Grid independence test*

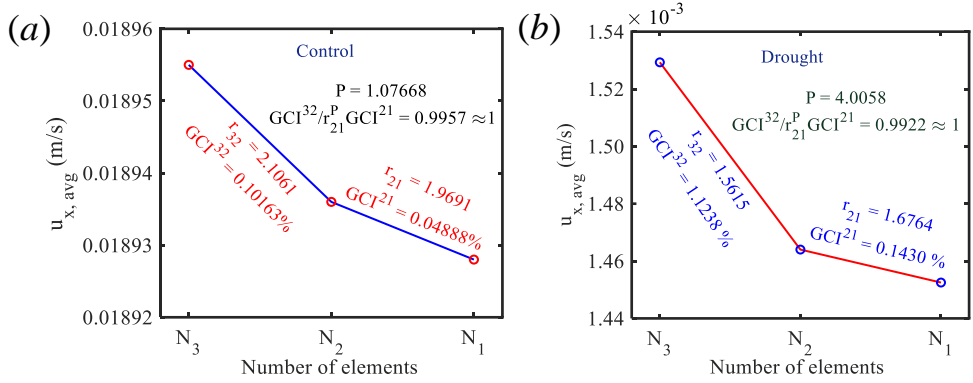

**Figure 7**. Comparison of average axial flow velocity calculated at a cross-section located at 112.5 μm from the inlet of the xylem vessel for different meshing elements employing grid convergence index for (*a*) control and (*b*) drought conditions.

To optimized the involved computational cost and time without sacrificing the accuracy of the numerical results, we conduct a grid independence test for both control and drought-stressed conditions. In control case, we consider the number of elements as $N_1$= 35619634, $N_2$ = 9186703 and $N_3$ = 2071030, while in the drought-stressed case, the elements are varied as $N_1$ = 9308360, $N_2$ = 3312169 and $N_3$ = 1358354. To this end, we appeal Richardson extrapolation method-based grid convergence index (GCI) (Celik *et al.* 2008) to estimate the error in the average axial velocity as a key variable. We calculate the average axial velocity at a cross-section located 112.5 μm from the inlet of the drought-affected xylem vessel. The calculated two grid convergence indices are $GCI_{fine}^{21}$ =0.04888% and $GCI_{fine}^{32}$ = 0.10163% for the control condition, whereas $GCI_{fine}^{21}$ =0.1430% and $GCI_{fine}^{32}$ = 1.1238% for the drought-stressed xylem vessel. The calculated indices lead to an asymptotic solution, yielding a convergence criterion $GCI_{fine}^{32} / r_{21}^{p} GCI_{fine}^{21} \approx 1$, as depicted in figures 7(*a*)-(*b*) for both the cases. Based on this analysis, we select N2 grids for all simulations performed in this study to ascertain that our numerical results are free from grid resolution bias.

*Axio-radial flow field, flow loading and mechanical stress*

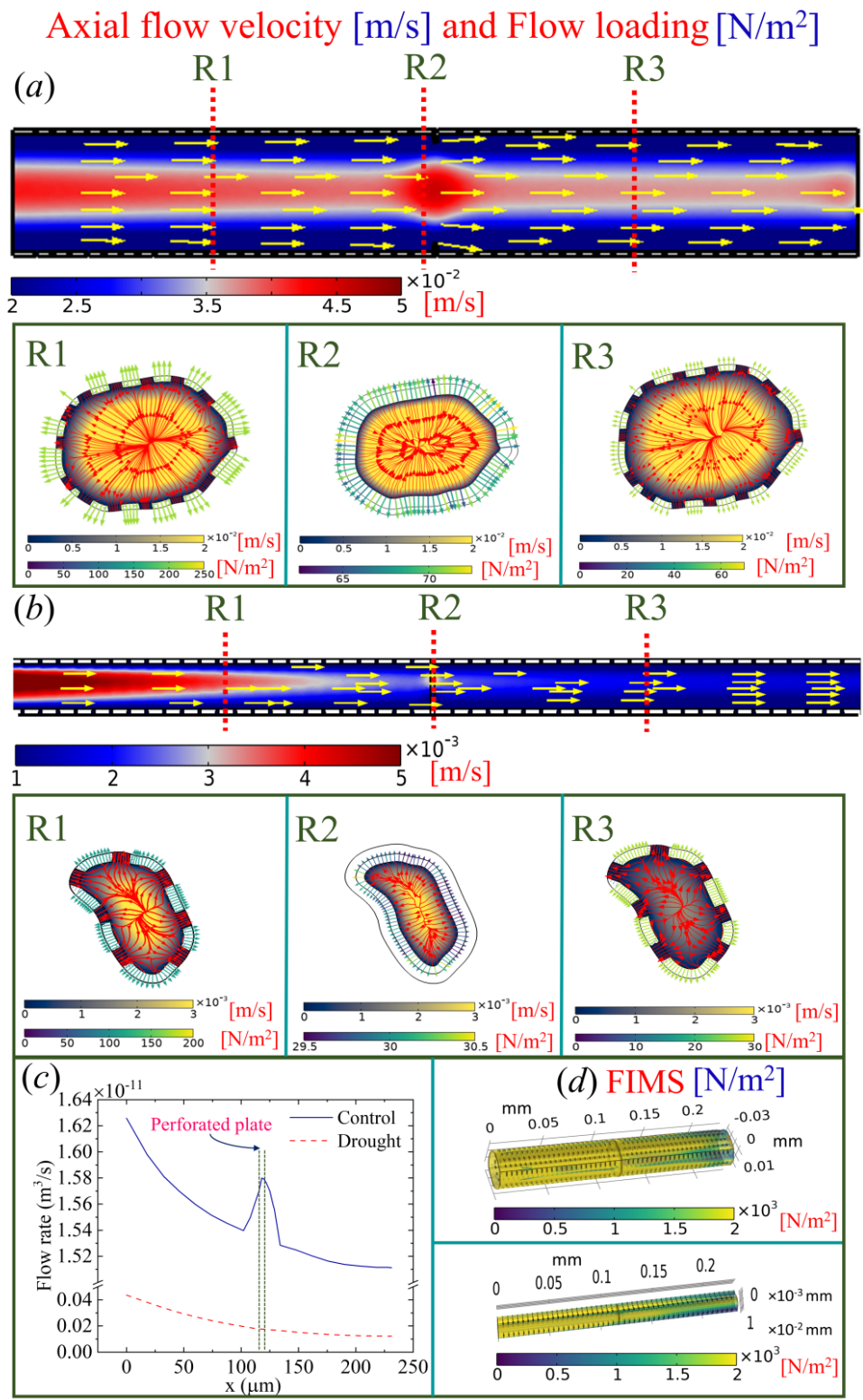

**Figure 8.** Contours of axial flow velocity along the length of xylem vessels as well as at three distinct lateral planes, R1, R2 and R3 located at x = 57.6 μm, 115.5 μm and 173.5 μm, respectively under (*a*) control and (*b*) drought conditions. (*c*) Axial flowrate profile along the xylem length for control and drought cases. (*d*) Contours of low-induced mechanical stress on the xylem wall under control and drought conditions.

We now analyze the local flow field developed within xylem vessels both in axial as well as radial directions. We undertake this effort to envision the effect of water availability at the xylem vessel through the demonstration of local flow patterns. Accordingly, we present in figures 8(a)-(b) the axial flow velocity contours with flow loading on the pitted wall and radial flow field in three distinct lateral planes (R1, R2 and R3) for plants under control and drought-stressed conditions respectively. Also, we show the flow rate through the xylem vessel and the variation of flow-loading induced mechanical stress (FIMS) acting on the inner wall surface of xylem vessel (also, known as lumen wall) in figures 8(c) and (d) respectively, for both control and drought conditions. It is evident from figures 8(*a*) and (*b*) that the axial flow intensity in the xylem vessels is greater for the plants grown-up under control condition as compared to the drought-stressed scenario. This observation is attributed to the existence of the larger pressure gradient in control conditions (cf. figures 5*a-b*). Notably, the intensified pressure gradient in control condition leads to a relatively higher flow velocity strength at the perforated plate (plane R2) due to the reduction in cross-sectional area therein as witnessed in figure 8(a). Moreover, the intensification of flow field near the perforated plate is not significantly evident in drought-stressed scenario, and this is because of weaker pressure gradient develops in the xylem vessel owing to water-deficient soil environment. As a result, the flow loading intensity at the lumen wall (hollow, central pathway within the xylem that facilitates transport of nutrient through the plant) is seen to be lower in drought-stressed scenario compared to the control conditions. It is interesting to note that, in region close to the perforated plate, the flow loading intensity is significantly higher for the control case (cf. figure 8a - plane R2) in comparison to the drought condition (cf. figure 8b - plane R2). This can be explained by the fact that, in control conditions, the local pressure near the upstream side of the perforated plate is stronger than that under drought-stressed conditions (cf. Appendix C). Additionally, from R1 to R3, the flow loading intensity decreases. This is due to a reduction in local pressure in the downstream (R1 to R3) region, which in turn, reduces flow-induced force at the lumen wall. Furthermore, the flow loading between R1 and R2 (up to the perforated plate) is significantly reduced due to the flow obstruction caused by the perforated plate. Whereas the absence of obstruction results in a moderate drop in flow loading between R2 and R3. The higher axial flow velocity and larger cross-sectional area in the control case result in greater axial throughput in the xylem vessels compared to the drought condition, as evident in figure 8(c). Moreover, the intensified flow velocity near the perforated plate region causes a sharp increase in flow rate in the control case. In contrast, this effect is absent in drought-stressed xylem vessels due to their reduced flow strength, as witnessed in the axial velocity profiles (cf. figures 8*a-b*).

We also show the flow loading-induced mechanical stress (FIMS) in figure 8 (*d*). It is observed that the FIMS is higher between R1 and R2 than that between R2 and R3. Furthermore, the increased flow loading on the lumen wall under control condition allows for a larger intensity of mechanical stress when compared to the drought-stressed scenario. Based on this discussion, we can deduce that increased flow loading at the lumen wall during control condition is sufficient for preserving the vessel's structural stability. Consequently, xylem vessel under control condition exhibits nearly circular shape (£→1) (cf. figure 2). In contrast, the decreased flow loading in drought-stressed scenario offers a relatively weak structural stability to the xylem vessel, and results in the deviation from the circular shape (£ > 1) (cf. figure 2). Quite interestingly, it is seen from figure 9(a)-(b) that the geometrical adaptations of xylem vessels to drought-stressed condition attempt to preserve a nearly identical radial flow intensity as in the control condition. *This observation is suggestive of that the architectural adaptations of xylem vessel to drought condition offer a self-sustaining mechanism to plant to supply adequate nourishment to the adjacent vessels in the prevailing situation with the compromisation of axial flow velocity, and increase the survivability of plants*. Also, the presence of pits in the R1 and R3 planes leads to the greater radial flow intensity as compared to the pit-less perforated plate region (R2). As the flow resistance increases towards the vessel ends, the difference in average radial flow velocity between control and drought-affected xylem vessels becomes negligible therein (cf. figure 9c). This indicates that reduction in axial flow velocity in drought-stressed scenario is compensated by the radial flow velocity through xylem pits, thereby contributing to plant survival.

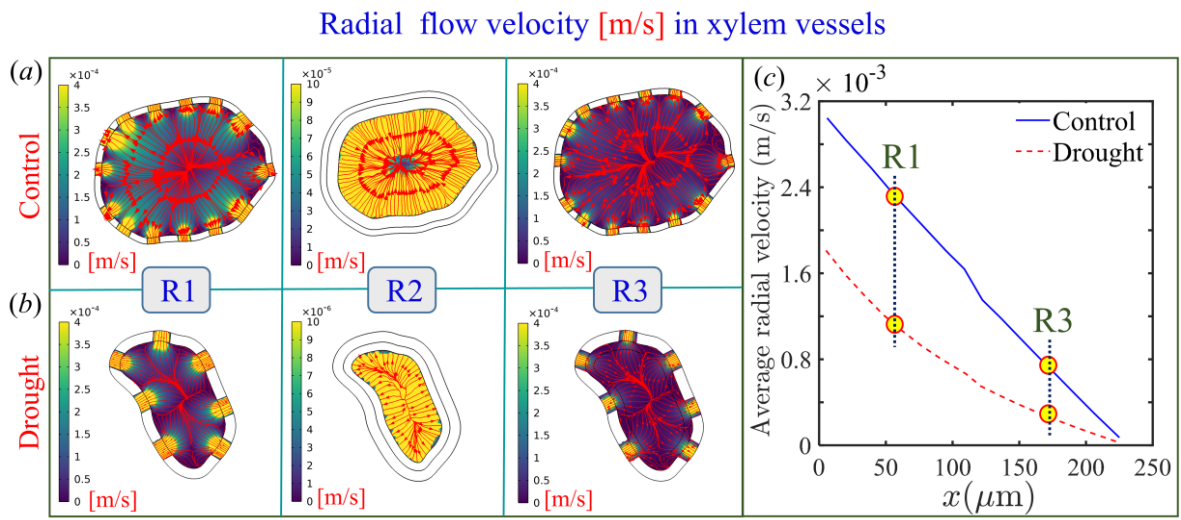

**Figure 9.** Contours of radial flow velocity at three distinct lateral planes namely R1 at x = 57.6 μm, R2 at x =115.5 μm and R3 at x =173.5 μm along the xylem vessels under (*a*) control and (*b*) drought conditions. (*c*) Average radial velocity in the xylem vessel under control and drought conditions.

Now, to examine the axial flow performance of the xylem vessels during control and drought conditions, we introduce specific hydraulic conductivity as (Quintana-Pulido *et al.* 2018):

$$K_h = \frac{\rho Q L}{\Delta p A_s} \quad (12)$$

Here, Q and $A_s$ are the axial flow rate and cross-sectional area of xylem vessels respectively. It may be added here that the effect of xylem pits on the radial transport is estimated using the radial flow efficiency of the xylem vessels under both control and drought conditions. This can be expressed mathematically as follows (Xu *et al.* 2020):

$$\eta_R = \frac{F_{R1} - F_{R2}}{F_{R1}} \quad (13)$$

Here, $F_{R1}$ and $F_{R2}$ represent the flow resistance in xylem vessels with and without (solid wall) pitted wall structures, respectively. The flow resistance in both cases is calculated using the relation: $F_{Rm} = \Delta p_m / Q_m$; where, $\Delta p_m$ and $Q_m = (\pi/4) D^2 u_{x,Avg,m}$ are the pressure drop across two sections of the xylem vessel and the corresponding flow rate at the downstream section, respectively. Here, $u_{x,Avg,m}$ denotes the average axial velocity, and the subscript $m =$ 1, 2 corresponds to xylem vessels with and without pit structures on the walls, respectively.

It is apparent from equation (13) that the radial efficiency of the xylem vessel is strongly dependent on the flow resistance ($F_{Rm}$). Therefore, the axial variation of flow resistance for both control and drought-stressed conditions is depicted in figures 10(*a*) and (*b*), respectively, considering pitted and solid wall models. For the control condition, the pitted wall model has mildly higher flow resistance as compared to the solid wall model, as seen from figure 10(*a*). We attribute this observation to the stronger axial pressure gradient available for the control case, which in turn, allows a relatively lower reduction in axial flow by radial leakage through the pitted wall, leading to higher flow resistance. Whereas the higher radial to axial average flow ratio in draught-stressed condition as compared to the control case (cf. figures 8 and 9) allows a relatively higher reduction in axial flow velocity. Despite this large decrease in axial flow rate in the drought-stressed case, the pitted wall model possesses a significantly higher flow resistance than the solid wall model, as shown in figure 10(*b*). This observation is attributed to the reduction in diameter of the xylem vessel under drought-stressed condition, since resistance to the axial flow is inversely dependent on the diameter of the flow pathways. The foregoing discussion suggests that axial flow velocity decreases (underscoring that $Q_m$ will drop) and resistance to flow axial flow increase (signifying, $\Delta p_m$ will drop) with shrinking of the xylem vessel's diameter due to drought stress. Following this reason, the xylem vessel under drought-stressed condition experience higher magnitude of axial flow resistance ($F_{Rm} = \Delta p_m/Q_m$) than the control case as the relative decrease in $Q_m$ is more than $\Delta p_m$.

Based on equation (12), the hydraulic conductivity of xylem is inversely proportional to the flow resistance ($F_R = \Delta p/Q$) of the pitted wall model. Consequently, the higher magnitude of flow resistance in the draught stress scenario allows lower hydraulic conductivity when compared to the control condition, as illustrated in figure 10(*c*). Furthermore, an increase in flow resistance in the axial direction reduces hydraulic conductivity under drought-stressed condition.

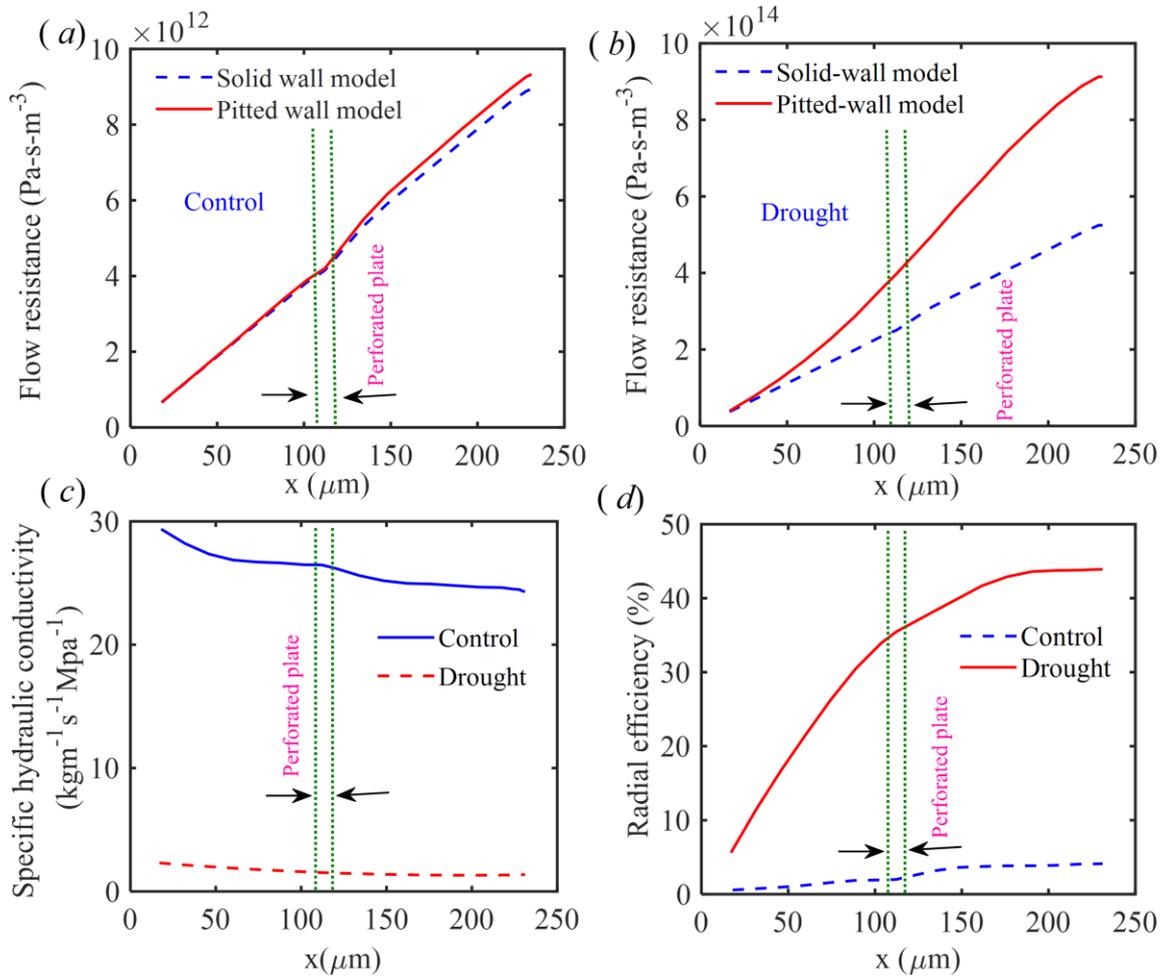

**Figure 10.** Axial variations in flow resistance for (*a*) control and (*b*) drought conditions. Axial variations in (*c*) specific hydraulic conductivity and (*d*) radial efficiency in the xylem vessels under control and drought environments.

Finally, figure 10(*d*) depicts the axial variation of radial flow efficiency under control and drought-stressed conditions. We observe that the significant difference in flow resistance between the pitted and solid wall models (see figure 10b) allows higher radial efficiency in drought-stressed situations as compared to the control condition. Based on these findings, we can infer that plants undergoing through drought-stressed situation adapt geometrical modulation of the xylem vessels to preserve radial transport of nutrients, aiding survival in water deficit environment by enhancing radial flow efficiency.

**Summary, perspective and outlook**

In this study, we resort to both experimental and numerical approaches to investigate how drought stress alters the morphological, biochemical, electrohydrodynamic, and structural traits of *B. juncea* (Indian mustard), and the influence of these variations on the local flow physics of its xylem. To this end, we conduct a series of experiments, including scanning electron microscopy, zeta potential measurements, inductively coupled plasma mass spectrometry analysis, laser Raman spectroscopy, atomic force microscopy, fluorescence microscopy and leaf chlorophyll content measurement. We also measure the streaming potential of xylem vessels to estimate the pressure gradient that drives the flow of nutrient solution through plants under both control and drought-stressed conditions. Our results show that drought stress significantly reduces xylem diameter and wall pit aperture size, signifying an adaptive mechanism to maintain hydraulic balance. The reduced water availability under drought-stressed scenarios lowers both streaming potential as well as pressure gradient inside the xylem vessels. Attributing to reduced cellulose content in drought-stressed condition, the cross-sectional shape of xylem vessels deviates from an ideal cylindrical shape, as seen in control condition, and also diminishes zeta potential and elasticity modulus. Drought also disrupts the plant's metallic element concentration, causing nutrient imbalances that impair essential bioactivities and compromise overall plant health. The significant alteration of key metabolites under drought-stressed conditions leads to increased reactive oxygen species and reduced chlorophyll content, which ultimately results in plant mortality. Also based on these experimental evidences, we develop a mathematical model of the xylem vessels corresponding to both control and drought-stressed conditions, and compute the flow and displacement fields employing three-dimensional numerical simulations. We reveal that axial flow field intensity becomes higher in the control case than in the drought-stressed scenarios, with a greater flow intensification in the perforated plate under control conditions. In drought conditions, the reduced flow loading significantly lowers the mechanical stress experienced by the lumen wall (hollow, central pathway within the xylem that facilitates transport of nutrient through the plant). Notably, drought-induced modifications in xylem vessel geometry help to maintain radial flow strength at levels comparable to the control case. Our results also show that, under drought stress, axial flow resistance is considerably higher when a pitted wall model is applied than with a solid wall model, leading to reduced hydraulic conductivity relative to the control scenario. Interestingly, radial flow efficiency is markedly elevated under drought stress, indicating that plants, being nature-inspired systems, adopt geometric adaptations to counteract water scarcity by enhancing radial flow efficiency and ensure optimal nutrient transport for their survival as well. Overall, our integrated experimental and numerical findings provide insight into how geometric adaptations in xylem vessels modify flow characteristics under drought-stressed conditions, ultimately supporting plant survival.

**Acknowledgements**

P.K.M. acknowledges the financial support provided by the SERB (DST), India, through Project No. CRG/2022/000762 from the SERB (DST), India. S.K.M. acknowledges the financial support provided by the National Post-Doctoral Fellowship (N-PDF) with Reference No. PDF/2023/002072.

**Declaration of Interests**

The authors report no conflict of interest.

**Data availability statement**

The data that support the findings of this study are available from the corresponding author upon reasonable request.

**Author contributions**

**Jinmay Kalita:** Formal analysis, Data curation, Conceptualization, Investigation, Methodology, Software, Experiments, Visualization, Validation, Writing – original draft.
**Sumit Kumar Mehta:** Formal analysis, Investigation, Methodology, Software, Validation, Experiments, Writing – original draft.
**Suraj Panja:** Formal analysis, Investigation, Methodology, Software, Experiments, Validation, Writing – original draft.
**Pranab Kumar Mondal:** Conceptualization, Funding acquisition, Methodology, Project administration, Resources, Supervision, Validation, Writing – review, and editing.

**Appendix A**. Reactive oxygen species (ROS) generation and leaf chlorophyll A content under control and drought conditions

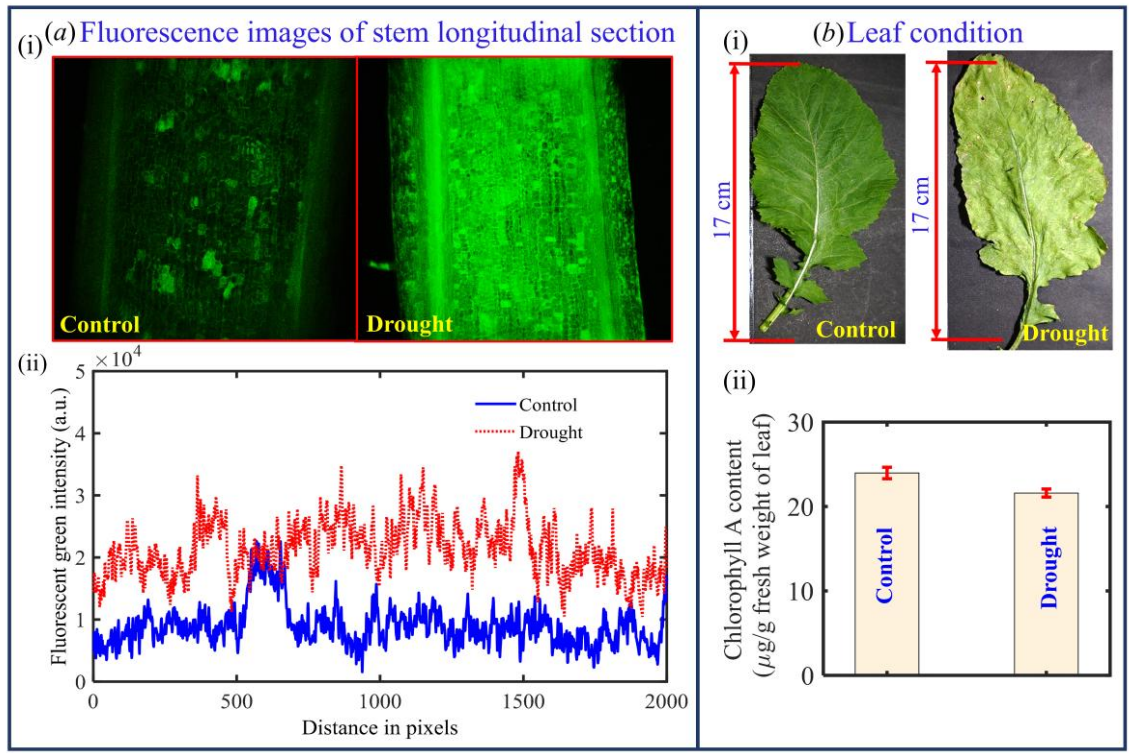

**Figure 11.** (*a*) Fluorescent microscopy images of a longitudinal section of the stem showcasing (i) qualitative and (ii) quantitative variations of ROS specific fluorescent intensity. (*b*) Condition of leaves under control and drought environments, highlighting (i) external morphological changes and (ii) variation in chlorophyll A contents.

In figure 11(*a*), we demonstrate the variation in ROS generation inside plant cells in control and drought-stressed plants alongside their quantitative variations obtained from fluorescent microscopy images. It is evident that the ROS generation in the plant cells under drought is significantly intensified. This is due to formation of free radicles such as, singlet oxygen ($^1O_2$), superoxide radical ($O^{2-}$), hydrogen peroxide ($H_2O_2$) and the hydroxyl radical (HO$^•$) (Hasanuzzaman *et al.* 2020). The enhancement of these molecules is attributed to the uneven distribution of phototoasssimilates from leaves to other organs of plants. The responsible element for this anomaly is deformation of chloroplast, which restricts the diffusion of $CO_2$ from the chloroplast membrane. Thus, there occurs a discrepancy between photoassimilates produced in the leaves and their utilization by plants. The non-utilized portion of photoassimilates ultimately triggers the ROS generation (Senbayram *et al.* 2015). We confirm this effect by illustrating the condition of leaves in control and drought-induced plants in figure 11(*b*). The leaves of drought-stressed plants are observed to be yellowish, which is termed as chlorosis (Ahanger *et al.* 2016) (cf. figure 13*b*ii). It is due to the reduced chlorophyll

A content in the leaves as shown in figure 11(*b*)(ii). This degradation in chlorophyll A level is mainly caused by excessive ROS generation inside the plant, triggered by magnesium deficiency in drought-stressed plants as discussed previously. The enhanced ROS production and reduced chlorophyll contents collectively reduce the photosynthetic activity during drought stress, adversely affecting the overall growth of the plant.

**Appendix B**. Analysis of streaming potential difference induced pressure drop for cylindrical micro-vessel

From the momentum transport equation, the reference velocity scale ($u_r$) in the dominating pressure-driven flow scenario through micro-vessel can be expressed as follows due to (White & Majdalani 2006; Clerx *et al.* 2020):

$$u_r \approx \frac{1}{\mu}\left(-\frac{\partial p}{\partial x}\right)R^2 \qquad \text{A(1.1)}$$

Now, Poisson's equation of the electrical-double layer (EDL) potential field ($\psi$) can be written as (Schnitzer *et al.* 2012):

$$\varepsilon_0 \varepsilon_r \nabla^2 \psi = -\rho_e \qquad \text{A(1.2)}$$

Here, $\rho_e$ is space charge density. From Eq. A(1.2), the scaled value of space charge density can be written in terms of zeta potential ($\zeta$) because of very thin EDL as the concentration of electrolyte is very high (cf. figure 5*b*) (Schnitzer *et al.* 2012). In a higher electrolyte concentration, the EDL thickness becomes $10^4$-$10^5$ times lesser than the radius of micro-vessel. Accordingly, from the order analysis, we have: $\varepsilon_0 \varepsilon_r \left(\frac{\zeta}{R^2}\right) \approx -\rho_e$. Hence, the order of streaming current ($I_s$), induced due to the pressure driven flow can be expressed as $I_s \approx \int \rho_e u_r dA$; here, $dA$ is the elemental cross-sectional area of the micro-vessel. Substituting the order value of space charge density, we can get $I_s \approx -\varepsilon_0 \varepsilon_r \left(\frac{\zeta}{R^2}\right) u_r A_s$; here, $A_s$ is the cross-sectional area of micro-vessel. Now, using Ohm's law, we can get streaming potential difference as $\Delta V_s = I_c \mathfrak{R}$. The conduction current, $I_c$, is neutralized by the streaming current, and we have $I_c = -I_s$. Here, electrical resistance ($\mathfrak{R}$) can be expressed in terms of electrical conductivity ($\sigma$), length ($L$) and cross-sectional area as $\mathfrak{R} = \frac{L}{\sigma A_s}$. Now, substituting the value of $I_c$ and $\mathfrak{R}$ in the expression of streaming potential difference, we can get: $\Delta V_s \approx \varepsilon_0 \varepsilon_r \left(\frac{\zeta}{R^2}\right) u_r A_s \left(\frac{L}{\sigma A_s}\right)$ or $\Delta V_s \approx \varepsilon_0 \varepsilon_r \left(\frac{\zeta}{R^2}\right)\left[\frac{1}{\mu}\left(-\frac{\partial p}{\partial x}\right)R^2\right] A_s \left(\frac{L}{\sigma A_s}\right)$. Using, pressure-difference as $\Delta p = L\left(\frac{-dp}{dx}\right)$, we have the final expression of streaming potential difference as:

$$\Delta V_s \approx \frac{\varepsilon_0 \varepsilon_r \zeta \Delta p}{\sigma \mu} \qquad \text{A(1.3)}$$

**Appendix C**. Pressure contours

Figure 12 shows the local pressure contours near to the perforated plate region. It is evident that the perforated plate allows the local pressure field alteration while obstructing the main flow. In contrast to the drought-stressed situation, the stronger axial flow permits a greater intensity of local pressure close to the upstream region of the perforated plate under control condition. This finding leads to the inference that the flow loading at the perforated plate under the control condition will be higher than in the drought situation.

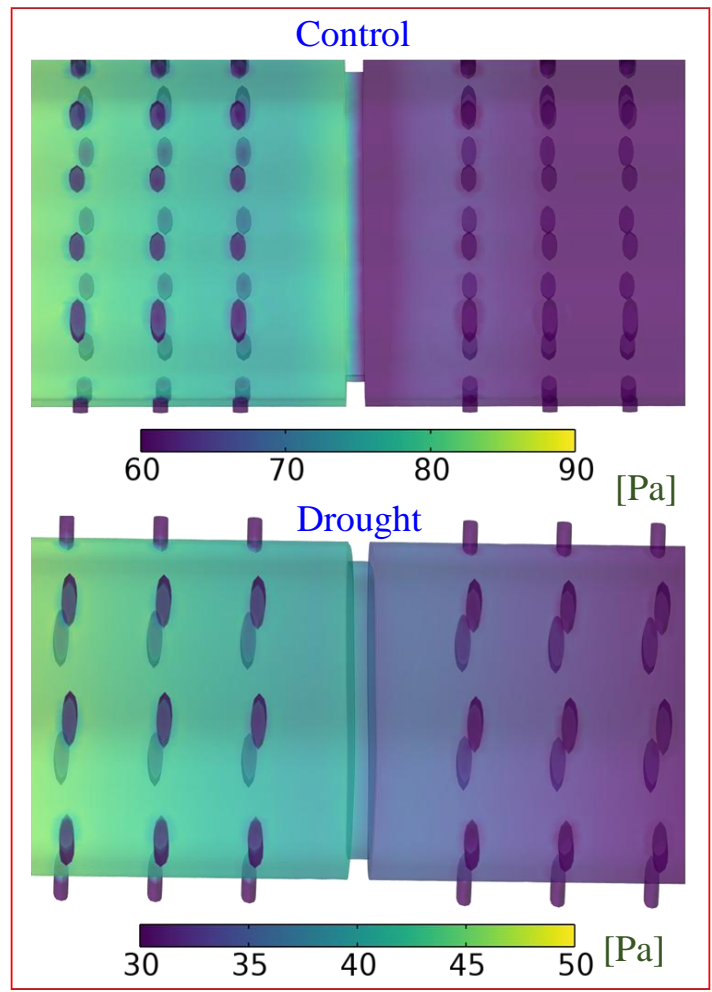

**Figure 12.** Contours of local pressure field near the perforated plate for control and drought condition.